\documentclass{aa}  

\usepackage{graphicx}
\usepackage{txfonts}

\usepackage{ulem}

\usepackage{xcolor}

\usepackage{soul}

\usepackage{float}

\usepackage{array, multirow, bigdelim, makecell, booktabs} 

\usepackage{threeparttable}

\usepackage{lscape}

\usepackage{longtable}

\usepackage{amssymb}

\usepackage[colorlinks = true, linkcolor=blue, citecolor = blue]{hyperref}
\begin{document}

\title{eROSITA-selection of new period-bounce Cataclysmic Variables 
}
\subtitle{First follow-up confirmation using TESS and SDSS}


   \author{Daniela Muñoz-Giraldo
          \inst{1}
          \and
          Beate Stelzer\inst{1}
          \and
          Axel Schwope\inst{2}
          \and
          Santiago Hernández-Díaz\inst{1}
          \and
          Scott F. Anderson\inst{3}
          \and
          Sebastian Demasi\inst{3}
          }

   \institute{Institut f\"ur Astronomie und Astrophysik, Eberhard Karls Universit\"at T\"ubingen, Sand 1, 72076 T\"ubingen, Germany \\   
   \email{munoz-giraldo@astro.uni-tuebingen.de} 
        \and
          Leibniz-Institut für Astrophysik Potsdam (AIP), An der Sternwarte 16, 14482 Potsdam, Germany 
        \and
          Department of Astronomy, University of Washington, Box 351580, Seattle, WA 98195, USA  \\    
             }

   \date{Received XX; accepted XX}

 
  \abstract
   {Between 40$\%$ and 80$\%$ of cataclysmic variables (CVs) are expected to have evolved past the period-minimum and contain a degenerate donor. However, observational surveys for CVs have only been able to detect a few of these highly evolved ``period-bouncers'', most likely due to the intrinsic faintness associated with their predicted low mass accretion rates.}
   {We searched for new period-bouncers in optical white dwarf (WD) catalogs. Our findings will establish if the missing population of these underrepresented type of CVs is hiding amongst these catalogs.}
   {We have produced an initial selection of high-likelihood period-bounce candidates based on our multiwavelength period-bouncer scorecard and selection cuts including X-ray data from the extended ROentgen Survey with an Imaging Telescope Array (eROSITA) on board the Spektrum-Roentgen-Gamma spacecraft (SRG). We have laid out a clear path based on three main requirements (classification as a CV, determination of an orbital period and detection of a very late-type donor) that should result in the confirmation of several of these candidates, and that will serve as a reference for future searches of period-bouncers.}
   {Our selection method to identify new period-bouncers in WD catalogs successfully retrieves already known period-bouncers found in the \textit{Gaia} DR3 WD catalog, identifying as well five new period-bouncers classified in the literature only as CVs, and $137$ high-likelihood period-bounce candidates that were classified in the literature only as single WDs. Our path for confirming these candidates has already produced its first successful result with the confirmation of GALEX J125751.4-283015 as a new period-bouncer. Several other candidates have already fulfilled at least one of our three requirements making their future confirmation likely.}
   {Our search for period-bouncers using the X-ray eROSITA emission of objects in optical WD catalogs has led to the confirmation of six new period-bouncers identified from the \textit{Gaia} DR3 WD catalog (five previously known CVs and one WD candidate), a 18$\%$ increase that brings the present population to $39$ systems. Both the selection method for period-bounce candidates and the confirmation path that we outlined will aid in future searches for new period-bounce candidates, contributing to the goal of resolving the discrepancy between the predicted high number of period-bouncers and the low number of these systems successfully observed to date.}

   \keywords{ stars: cataclysmic variables - X-rays: binaries, cataclysmic variables - X-rays: surveys }

   \maketitle
%

\section{Introduction}\label{sect:intro}


Cataclysmic variables (CVs) are close binary systems in which a white dwarf (WD) accretes material from a late-type companion star that fills its Roche lobe \citep{warner2003}. As described by \cite{paczynski1976}, CVs form as a result of a common envelope (CE) phase during the binary’s evolution. In this phase, the envelope of a more massive, Roche-lobe filling primary star expands and engulfs its companion. After the ejection of the envelope, a post-CE binary is left consisting of the evolved core of the primary star (now a WD) and a low-mass, main-sequence companion. The friction generated during the CE phase causes a loss of angular momentum and energy, which drives a significant decrease in the orbital separation of the binary, such that, once it becomes small enough to permit mass transfer from the companion to the WD, the system morphs into a CV.

The subsequent evolution of CVs is driven by angular momentum losses towards increasingly shorter separations (and thus orbital periods $P_{\rm orb}$), resulting in a evolution track from longer orbital periods to shorter ones (\cite{paczynski1976}, \citealt{kolb1993}, \citealt{warner2003}). Starting as a CV with long orbital periods (P$_{\rm orb}>$ 3h), the system will evolve until it reaches the period gap (2h $<$ P$_{\rm orb}<\,$3h) becoming a detached binary. The system re-emerges at the lower end of the period gap as an active CV when the donor is once again filling its Roche lobe \citep{kolb1998}. The evolution continues toward even shorter periods through angular momentum loss due to gravitational radiation, until the system reaches a period minimum located at $P_{\rm min} \approx 80$ \,min \citep[see e.g.,][]{howell2001,patterson2011,pretorius2013,goliasch,mcallister2019,pala2020,pala2022}. At this point, the donor is out of thermal equilibrium due to its mass-loss timescale becoming much shorter than its thermal timescale, causing the donor to stop shrinking in response to mass loss \citep{king1988}. The very late-type donor, not being able to sustain hydrogen burning, becomes a brown dwarf \citep{howell2001}, with this change in internal structure resulting in the increase of the system's orbital separation and consequently in the CV bouncing back to longer orbital periods. The systems that go through this process of "bounce-back" to longer periods are known as period-bouncers \citep{patterson1998}.

\begin{table*}
    \begin{threeparttable}
    \centering
    \caption{
    $33$ confirmed period-bouncers including $21$ bona-fide period-bouncers. This is a unified list from \cite{munoz2024} and \cite{munoz2024b}.}
    \label{confirmedPB}
    \begin{tabular}{lcccccc} 
    \hline
    \noalign{\smallskip}
     & &  & \multicolumn{2}{c}{Score [$\%$]} & & \\
    \cmidrule(lr){4-5}
    Period-bouncer & Donor SpT & Reference & Full & Reduced & P$_{\rm WD}$ & eROSITA detection\\
    \noalign{\smallskip}
    \hline
    \noalign{\smallskip}
    V379 Vir & L8 (S) & \cite{farihi2008} & 100 & 100 & 0.996 & x \\
    SDSS J15141+0744 & L3 (P) & \cite{breedt} & 69 & 83 & 0.778 & \\
    PM J12507+1549 & M8 (P) & \cite{breedt} & 67 & 100 & 0.979 & x\\
    SDSS J10575+2759 & L5 (P) & \cite{mcallister2017} & 85 & 80 & 0.872 & x\\
    SDSS J14331+1011 & L1 (S) & \cite{santisteban2016} & 75 & 61 & 0.951 & \\
    WZ Sge & L2 (S)& \cite{harrison2015} & 90 & 100 & 0.986 & \\
    SDSS J10353+0551 & L0 (P) & \cite{southworth2006} & 88 & 87 & 0.996 & x\\
    SMSS J16063-1000 & L2 (P) & \cite{kawka2021} & 70 & 100 & 0.992 & \\
    QZ Lib & T (S) & \cite{pala2018} & 100 & 100 & 0.979 & x\\
    GD 552 & L0 (S) & \cite{unda2008} & 79 & 67 & 0.959 & \\
    MT Com & L (P) & \cite{patterson2005} & 76 & 100 & 0.971 & \\
    V455 And & L2 (P) & \cite{araujo2005} & 88 & 92 & 0.637 &\\
    V406 Vir & L3 (S) & \cite{pala2019} & 64 & 72 & 0.979 & x\\
    BW Scl & T (S) & \cite{neustroev2022} & 82 & 80 & 0.960 & x\\
    EZ Lyn & L2 (P) & \cite{zharikov2013} & 93 & 100 & 0.991 & \\
    CRTS J12222-3115 & L0 (P) & \cite{neustroev2017} & 77 & 75 & 0.976 & x\\
    V1108 Her & L1 (S) & \cite{ishioka2007} & 81 & 67 &  & \\
    GALEX J04113+6853 & T (S) & \cite{galiullin2024} & 65 & 75 & 0.986 & - \\
    HV Vir & L5 (P) & \cite{mennickent2002} & 61 & 73 & 0.926 & x\\
    WD J18204-0422 & T5 (P) & \cite{cunningham2025} & 79 & 100 & 0.991 &  \\
    WD J19070+2052 & T6 (P) & \cite{cunningham2025} & 79 & 100 & 0.991 &  \\
    LP 731-60 &  &  & 92 & 87 & 0.995 & x\\
    EG Cnc &  &  & 89 & 83 & 0.995 & x\\
    SDSS J12160+0520 &  &  & 100 & 100 & 0.984 & x\\
    1RXS J02323-3718 &  &  & 67 & 83 & 0.984 & x\\
    WISE J11121-3538 &  &  & 94 & 89 & 0.839 & x\\
    PM J12192+2049 &  &  & 81 & 87 & 0.990 & x\\
    PNV J17144-2943 &  &  & 80  & 67 & 0.988 & x\\
    CRTS J10441+2113 &  &  & 67  & 73 & 0.965 & x\\
    SDSS J07550+1435 &  &  & 62 & 80 & 0.994 & x\\
    1RXS J10142+0638 &  &  & 92 & 100 & 0.986 & x\\
    2QZ J14283+0031 &  &  & 72 & 72 & 0.985 & x\\
    eRASS J05472+1326 & & & 100 & 100 & 0.984 & x\\
    \hline
    \end{tabular}
    \begin{tablenotes}
        \small
        \item We include the confirmation of a late-type donor in the system through spectroscopy (S) or photometry (P) for $21$ confirmed period-bouncers (referred to as ``bona-fide'' period-bouncers). The remaining $12$ confirmed period-bouncers do not yet have a detected late-type donor. The score columns refer to the final value of each system according to: the period-bounce scorecard defined by \cite{munoz2024} and the reduced period-bounce scorecard (see Sect.~\ref{subsect:reducedScorecard}). The probability of being a WD is presented for the period-bouncers found in \cite{gentile2021}. We indicate in the last column the period-bouncers with an eROSITA detection from the western ($x$) and eastern (-) Galactic half-sky. 
    \end{tablenotes}
  \end{threeparttable}
\end{table*}

The majority of the CV population, according to theoretical models describing CV evolution, is expected to be made-up of period-bouncers. However, the actual contribution from period-bouncers to the total CV population is highly disputed, with estimations ranging between 40$\%$ and 80$\%$, depending heavily on assumptions about the formation, evolution and system parameters (see e.g., \citealt{kolb1993}; \citealt{goliasch}; \citealt{belloni2020}). To this date, there are $33$ confirmed period-bouncers, including $21$ bona-fide period-bouncers which have a reliable detection of a late-type donor that strengthens their classification (see Table~\ref{confirmedPB}). Estimates place the percentage of observed period-bouncers between 3$\%$ and 25$\%$ (\citealt{inight2023a}, \citealt{pala2020}, \citealt{rodriguez2025}), establishing these objects as a minority subtype of CVs. 

The late-type donors (late M to T dwarfs) of period-bouncers are characterized for presenting emission in bands like X-ray (\citealt{audard2007}, \citealt{deLuca2020}) and optical (\citealt{girven2011}, \citealt{inight2023b}, \citealt{inight2023a}) that is hardly detectable with present-day instrumentation, where the emission is dominated by the accretion structures and WD in each case respectively. Therefore, the detection of X-ray emission arising from an optically identified WD is a key diagnostic of ongoing mass accretion in a WD-dominated binary system, and hence it is the most promising path for the identification of period-bouncers. 

Optical surveys like \textit{Gaia} or Sloan Digital Sky Survey (SDSS) have produced vast catalogs of WDs \citep[see e.g.,][]{kepler2016,jimenez2018,gentile2019,gentile2021,kepler2021}. These catalogs can be used as a starting point for the search of the missing population of period-bouncers as demonstrated in \cite{inight2023a}. Particularly, we are searching for emission detected by an all-sky X-ray survey that can prove the systems in these catalogs are not single cool WDs but potential interacting binaries. The launch of the extended ROentgen Survey with an Imaging Telescope Array \citep[eROSITA;][]{predehl2021} on board the Spektrum-Roentgen-Gamma mission \citep[SRG;][]{sunyaev2021} has allowed us access to large statistical samples of X-ray sources. The eROSITA detections of confirmed period-bouncers (\citealt{munoz2023}, \citealt{munoz2024}, \citealt{munoz2024b}) proved the capabilities of this instrument in the identification and confirmation of such faint sources, leading to an increase of 82$\%$ in the population of eROSITA-detected period-bouncers. 

The main focus of this study is to find the population of period-bouncers that is potentially hiding in WD catalogs. To this end, we have used the selection cuts and period-bouncer scorecard by \cite{munoz2024} to identify high-likelihood period-bounce candidates, as well as other interesting X-ray sources like single WDs, wide WD-main sequence binaries, WD-WD binaries, ultracompact binaries, amongst others. In our effort to confirm the period-bounce candidates we have established three main requirements that a candidate needs to fulfill in order to be successfully reclassified from a single WD into a period-bouncer. Observational follow-up campaigns are already on their way to ensure we have the data necessary to determine if our candidates fulfill these requirements. As more candidates are confirmed as period-bouncers, it will also provide a reliable observational data set that could be used for any future theoretical modelling of the late-phase of CV evolution.

Our selection of period-bounce candidates is introduced in more detail in Sect.~\ref{sect:selection}, providing specific information on the WD catalog and scorecard used for this purpose. In Sect.~\ref{sect:erosita} we discuss the eROSITA X-ray detections of the selected candidates which lead to the identification of high-likelihood period-bounce candidates. A breakdown of these high-likelihood period-bounce candidates is presented in Sect.~\ref{sect:PBcandidates}, together with a description of the necessary requirements for a confirmation, and the first successful re-classification of candidate GALEX\,J125751.4-283015 as a new period-bouncer. We present our conclusions in Sect.~\ref{sect:conclusions}.

\section{Selection of period-bounce candidates}\label{sect:selection}

In our series of papers aiming at the confirmation and characterization of new period-bouncers (\citealt{munoz2023}; \citealt{munoz2024}; \citealt{munoz2024b}) we have detected $9$ known period-bouncers using eROSITA and confirmed $12$ new period-bouncers through multiwavelength analysis also involving eROSITA ($6$ of the new period-bouncers had already been suggested as candidates in the literature). This represents a 33$\%$ increase in the population of confirmed period-bouncers, proving the usefulness of X-ray data, specifically from eROSITA, in the search for the missing population of period-bouncers.

Continuing our effort in the identification of new period-bouncers, we now search for X-ray emission from a population of \textit{Gaia}-selected WD candidates \citep{gentile2021}. We then apply a reduced version of the multiwavelength period-bouncer scorecard introduced in \cite{munoz2024} to those \textit{Gaia} candidate WDs with a reliable eROSITA X-ray detection. This way we define high-likelihood period-bounce candidates that have multiwavelength characteristics resembling those of confirmed members of this class.

\subsection{The \textit{Gaia} WD candidate catalog by \cite{gentile2021}}\label{subsect:WDcatalog}

We use the \textit{Gaia} Data Release 3 (DR3) catalog of 1280266 WD candidates by \cite{gentile2021} as a starting point for the search of new period-bouncers. This is because in the optical band period-bouncers are very likely to appear like single WDs, specially if they are in a quiescent state, since the contribution of the highly evolved secondary in period-bouncers is expected to be negligible at optical wavelengths \citep{santisteban2018}.

The \cite{gentile2021} WD catalog was compiled using \textit{Gaia} data for spectroscopically confirmed SDSS Data Release 16 WDs to construct broad cuts in the Hertzsprung–Russell (H-R) diagram that span the entire parameter space occupied by WDs. Further selection cuts based on absolute magnitude, color, and \textit{Gaia} quality flags were used by \cite{gentile2021} to remove the majority of contaminating objects. Fits of standard hydrogen atmosphere spectral models \citep{tremblay2011} to the \textit{Gaia} data were then used to obtain several parameters of the WD, including effective temperature, surface gravity and mass. As the effective temperature of the WD increases, \textit{Gaia} colors become less sensitive to it \citep{carrasco2014}. However, the WDs in period-bouncers are cool \citep{pala2022}, such that the effective temperature values obtained by \cite{gentile2021} from \textit{Gaia} colors should be accurate (see detailed discussion in Sect.~\ref{subsect:reducedScorecard}). 

\begin{figure}
    \centering
    \includegraphics[width=\columnwidth]{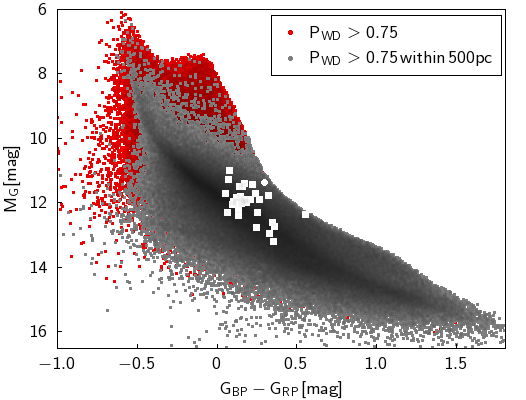}
    \caption{Population of ``high-fidelity'' WDs from \cite{gentile2021} with the confirmed period-bouncers in this population (white squares). The only confirmed period-bouncer with a distance larger than 500pc is highlighted as a circle.}
    \label{fig:GF_PB}
\end{figure}

For the selection of period-bounce candidates, the most relevant parameter from the \cite{gentile2021} catalog is the probability of being a WD (P$_{\textrm{WD}}$). This parameter was calculated by \cite{gentile2021} for each member of the catalog using as reference a sample of 22998 spectroscopically confirmed WDs from SDSS and 7124 contaminants (mainly composed of stars and Quasi-Stellar Objects). They found that about 91$\%$ of the spectroscopically confirmed WDs from SDSS are recovered when selecting objects with P$_{\textrm{WD}} >$\,0.75 \citep{gentile2021}. Our match of these ``high-fidelity'' WDs (P$_{\textrm{WD}} >$\,0.75) with the list of confirmed period-bouncers reported in Table \ref{confirmedPB} results in 94$\%$ of them being recovered. These ``high-fidelity'' WDs are therefore an ideal starting sample for a search of new period-bouncers. In Fig.~\ref{fig:GF_PB} we show the ``high-fidelity'' WD sample from \cite{gentile2021} that populates the WD locus and area surrounding it, and we highlight the $31$ confirmed period-bouncers included in this sample.

The search for period-bouncers, which are known for exhibiting very low mass accretion rates and therefore constituting the faintest CVs, should be distance-limited in order to be effective. Motivated by the eRASS sensitivity limit and the reduced accuracy in {\it Gaia} distance measurements achieved for faint objects (see \citealt{munoz2024} for more details), we enforce a distance limit of 500pc for the ``high-fidelity'' WDs (see Fig.~\ref{fig:GF_PB}) thus narrowing down the sample to 246731 \textit{Gaia} objects that resemble spectroscopically confirmed WDs in the optical band.

\subsection{A reduced version of the period-bouncer scorecard}\label{subsect:reducedScorecard}

Our period-bouncer scorecard, first introduced by \cite{munoz2024}, is composed of ten parameters that were constructed following defined characteristics for an ideal period-bouncer found in the literature: spectral type of the donor (\citealt{knigge2011}, \citealt{kirkpatrick1994}), donor mass \citep{knigge2011}, orbital period \citep{gansicke2009}, WD temperature \citep{pala2022}, {\it Gaia} variability \citep{inight2023b}, {\it Gaia} colors (\citealt{jimenez2018}, \citealt{gentile2021}), SDSS colors \citep{inight2023b}, ultraviolet colors \citep{patterson2011}, infrared colors \citep{littlefair2003}, and infrared excess (\citealt{girven2011}, \citealt{owens2023}). The application of the period-bouncer scorecard to our catalog of known CVs around the period-bounce made up of $217$ systems proved the usefulness of this tool as it produced a list of $76$ high-likelihood period-bouncer candidates of which $12$ were successfully confirmed (\citealt{munoz2024}, \citealt{munoz2024b}) thanks to the use of two additional X-ray parameters based on the eROSITA X-ray luminosity and X-ray-to-optical flux ratio of ``bona-fide'' period-bouncers.

\begin{table*}
    \centering 
    \caption{Completeness and accuracy of the parameters in the period-bouncer scorecard defined in \cite{munoz2024} when applied to a population of $219$ known CVs around the period-bounce. The numbers represent percentages.}             
    \label{CompAcc}                             
    \begin{tabular}{lcccccccccc}         
    \hline              
    \noalign{\smallskip}
    & \multicolumn{2}{c}{Donor} & & & & \multicolumn{4}{c}{Color} & \\
    \cmidrule(lr){2-3}\cmidrule(lr){7-10}
    & SpT & Mass & Orbital period & WD temperature & \textit{Gaia} variability & Gaia & SDSS & UV & IR & IR excess \\
    \noalign{\smallskip}
    \hline                      
    \noalign{\smallskip}
    Completeness & 43 & 56 & 57 & 70 & 97 & 84 & 81 & 43 & 53 & 93\\
    Accuracy & 90 & 25 & 25 & 37 & 29 & 47 & 21 & 53 & 45 & 25 \\
    \hline                                  
\end{tabular}
\end{table*}


The scores of the period-bouncers allows us to establish the performance of each parameter in the scorecard using its ``completeness'' and ``accuracy''. Here, ``completeness'' refers to the percentage of confirmed period-bouncers (Table \ref{confirmedPB}) that obtain the highest score for the parameter, quantifying how many confirmed period-bouncers resemble characteristics of an ideal period-bouncer from the literature. ``Accuracy'' refers to the percentage of systems with the highest score for the parameter which are confirmed period-bouncers, measuring the ``contamination'' from non-period-bouncers. The values for ``completeness'' and ``accuracy'' are given in Table \ref{CompAcc} for all $10$ parameters from the scorecard. 

Overall, the values for the ``completeness'' show that the characteristics for an ideal period-bouncer from the literature describe around half or more of the actual population of confirmed period-bouncers. The ``accuracy'', on the other hand, shows that the characteristics for an ideal period-bouncer from the literature (with the exception of the spectral type of the donor) also apply to a considerable number of systems that have not been confirmed as period-bouncers. This suggests that there are systems amongst the ``contaminants'' which might be part of the missing population of period-bouncers.

The parameter with the highest ``accuracy'' is the ``spectral type of the donor''. This is not unexpected, considering that a spectroscopic detection of a late-type donor in a CV is the ultimate confirmation that the system is a period-bouncer. However, the ``completeness'' of this parameter is only around 40$\%$ mainly due to the lack of a spectroscopic confirmation of the very late-type donor in several confirmed period-bouncers (see Table \ref{confirmedPB}). ``Ultraviolet colors'' is the only other parameter with an ``accuracy'' higher than 50$\%$, with 53$\%$ of systems that fulfill this selection cut being confirmed period-bouncers. The parameter that performs the best overall is ``\textit{Gaia} colors'', as it manages to retrieve 83$\%$ of confirmed period-bouncers while also selecting amongst the least amount of ``contaminants''. According to this, it is likely that a CV with \textit{Gaia} colors locating it within the selection cuts presented in the scorecard, is indeed a period-bouncer. This is specially useful considering that \textit{Gaia} colors are easily available (retrieved for 76$\%$ of our known CVs around the period-bounce), which is not the case for the ``spectral type of the donor'' (retrieved for 20$\%$ of our known CVs around the period-bounce).  

Considering that they are able to recover the majority of already confirmed period-bouncers (``completeness'' in Table~\ref{CompAcc}), while at the same time accurately identifying candidates that are actual period-bouncers (``accuracy'' in Table~\ref{CompAcc}), the most useful and relevant parameters when deciding if a CV is a period-bouncer are: spectral type of the donor, \textit{Gaia} colors, and ultraviolet colors.

Because the scorecard will be applied to a population of objects that have been published in the \cite{gentile2021} WD catalog as single objects, we do not have available information for the first three parameters (spectral type of donor, donor mass and orbital period) for a significant number of objects, as they correspond to parameters relating to the system being a binary. Therefore, we define a reduced version of the period-bouncer scorecard that is composed of the remaining seven parameters which mostly rely on multiwavelength photometry: WD temperature, \textit{Gaia} variability, \textit{Gaia} colors, SDSS colors, UV colors, IR colors and IR excess. 

\begin{figure*}
    \centering
    \includegraphics[width=0.49\textwidth]{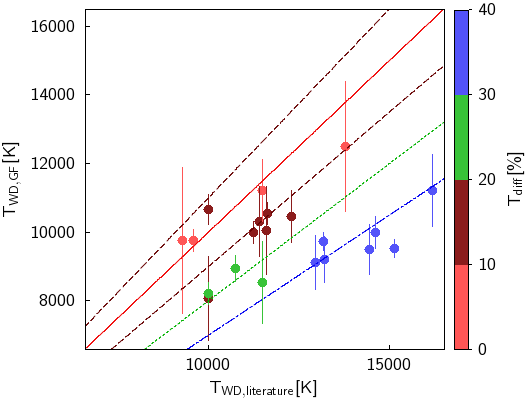}\hfill
    \includegraphics[width=0.49\textwidth]{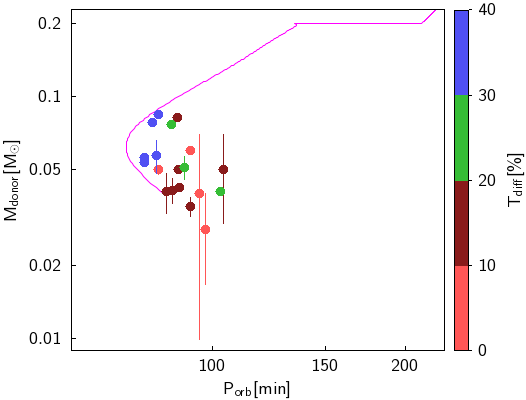}
    \caption{\textit{Left panel:} Comparison between WD temperatures obtained from individual literature studies versus those of the \cite{gentile2021} WD catalog. Values from the WD catalog that agree within the uncertainties with the literature values are shown in light red. We also show differences between these values of less than 10$\%$ (in brown), between 10$\%$ and 20$\%$ (in green), and larger than 20$\%$ (in blue). \textit{Right panel:} Orbital period versus donor mass. Standard evolution track for CVs \citep{knigge2011} is shown in pink as a reference.}
    \label{fig:Tdiff}
\end{figure*}

For the reduced version of the period-bounce scorecard we will exclusively use the WD temperature  reported in the \cite{gentile2021} catalog derived from \textit{Gaia} photometry, assuming that it represents the WD temperature of a period-bouncer in quiescence and not in an outburst. Out of the confirmed period-bouncers found amongst the ``high-fidelity'' WDs, $21$ have reliable WD temperatures in quiescence determined previously in the literature through a detailed study, which can be compared to the their respective WD temperature reported by \cite{gentile2021}. As can be seen Fig.~\ref{fig:Tdiff} (left panel), $14$ of $21$ confirmed period-bouncers have a WD temperature reported by \cite{gentile2021} within 20$\%$ of the value reported previously in the literature. This figure also shows that the sensitivity of \textit{Gaia} color to WD temperature decreases as the latter increases \citep{carrasco2014}, with the reliability of \textit{Gaia} derived WD temperatures falling particularly for values greater than 12500K. Additionally, the confirmed period-bouncers with the lowest difference between their WD temperature values are located mainly in the area after the period-bounce (see Fig.~\ref{fig:Tdiff} right panel), which indicates that using the WD temperature reported in the \cite{gentile2021} WD catalog is reliable specially for evolved period-bouncers located after the point of the period-bounce. 


The final score was then calculated considering only the seven parameters from the reduced scorecard. We are confident that the results from the reduced scorecard are reliable as it includes two of the three best performing parameters (\textit{Gaia} and ultraviolet colors) which, besides from the spectral type of the donor, make them the most relevant parameters for the identification of period-bouncers (see Table~\ref{CompAcc} and discussion above).

In Table~\ref{confirmedPB}, we show the final reduced scores for the confirmed period-bouncers calculated using the value reported by \cite{gentile2021} for the WD temperature. We see three tendencies among the confirmed period-bouncers: (1) Systems for which the score does not change significantly when using the full or reduced scorecard, indicating that both the multiwavelength photometry and the binary parameters obtained for them are consistent with a classification as a period-bouncer; (2) Systems for which the reduced score increases compared to the full score, meaning that these systems have binary parameters in the pre-bounce area inconsistent with the period-bouncer status suggested by their multiwavelength photometry. At least two of these systems have uncertain donor mass determinations that cause this lower full value for the score (SDSS J151415.65+074446.4 and PM J12507+1549; \citealt{breedt}); (3) Systems for which the reduced scores decreases compared to the full score. These are systems with multiwavelength photometry that falls outside the expectations for period-bouncers, and the strongest reason to be classified as period-bouncers is their securely determined binary parameters.


\section{eROSITA data}\label{sect:erosita}

Between December 2019 and December 2021 eROSITA has carried out four full-sky surveys, correspondingly named eRASS\,1 to eRASS\,4 \citep{merloni2024}. Source catalogs from eRASS data are produced at Max Planck Institut für extraterrestrische Physik (MPE) in Garching, Germany, with the eROSITA Science Analysis Software System (eSASS) described by \cite{brunner2022}. These catalogs comprise all eRASS sources in the western half of the sky in terms of Galactic coordinates (Galactic longitude $l \geq 180^\circ$), which is the sky area with German data rights on the eROSITA observations. From the list of $246731$ ``high-fidelity'' WDs within 500pc, $124657$ are located in the area with German data rights. 

To obtain the highest sensitivity for detecting the presumably faint period-bouncers hiding in the \cite{gentile2021} WD catalog, we used the merged catalog eRASS:4 which was generated from summing data from the first four all-sky surveys. The version of the eRASS:4 catalog used was produced with the data processing version 020\footnote{The source catalog used in our work is\\   all{\textunderscore}s4{\textunderscore}SourceCat3B{\textunderscore}221031{\textunderscore}poscorr{\textunderscore}mpe{\textunderscore}clean.fits (for eRASS:4).}. Source detection was performed in this catalog for three energy bands: $0.2-0.6$\,keV (soft band), $0.6-2.3$\,keV (medium band) and $2.3-5.0$\,keV (hard band).

\subsection{\textit{Gaia} WDs with eROSITA counterparts}\label{subsect:erositaCounterparts}

We corrected the coordinates of the \cite{gentile2021} WD catalog to the mean observing date of eRASS:4 using the individual {\it Gaia} DR3 proper motions of the systems. We then used the corrected coordinates to produce a match with the eRASS:4 catalog, allowing for a maximum separation of sep$_{\rm ox}=$ 30$^{\prime\prime}$, and enforcing the condition sep$_{\rm ox} < 3 \times RADEC\_ERR$, where $RADEC\_ERR$ is the positional error of the X-ray coordinates in units of arcseconds. We also checked that the eROSITA counterpart corresponds to a point source ($EXT = 0$ in the eRASS catalogs). This way we found that $912$ of the $124657$ ``high-fidelity'' WDs within 500pc located in the area with German data rights are detected in the eRASS:4 catalog, referred to from this point on as the ``eROSITA WD subsample''. Performing a reverse match between the eRASS:4 counterpart and the \textit{Gaia} DR3 catalog, allowing for a maximum separation of $30^{\prime\prime}$, we obtain that $601$ members of the ``eROSITA WD subsample'' where the closest \textit{Gaia} DR3 source to the eRASS:4 source corresponds to the \textit{Gaia} DR3 WD from \cite{gentile2021}, and are therefore considered to have a reliable counterpart. This includes $122$ possible resolved binaries due to a similar \textit{Gaia} parallaxes and proper motions ($24$ are known WD-MS wide binaries and $4$ are known WD-WD wide binaries from \citealt{elBadry2018}). The remaining $311$ members have a \textit{Gaia} DR3 source closer to the eRASS:4 source different from the \textit{Gaia} DR3 WD from \cite{gentile2021} that we have assigned as a counterpart, and should be examined in a case-by-case basis.


The ``eROSITA WD subsample'' with a reliable counterpart, thus, makes for only 0.48$\%$ of the ``high-fidelity'' WDs within 500pc. This is most likely because the majority of the undetected objects are indeed single WDs that produce faint X-rays below the eROSITA sensitivity limit or no X-rays at all. 

The eROSITA catalog holds the flux for a power law model with an index $\Gamma = 2.0 $ and a galactic absorption of $N_{\rm H} = 3 \times 10^{20} \rm cm^{-2}$ \citep{brunner2022}. Since this spectral model is not appropriate for CVs, which are characterized by a thermal plasma, we computed fluxes for an {\sc apec} model, following the procedure detailed by \cite{munoz2023}, with the help of a simulation that provided us with the eROSITA conversion factor, from count rate to thermal flux. We used an {\sc apec} model with $kT = 2.62\,$keV, $N_{\rm H} = 2.3 \times 10^{20}\, \rm cm^{-2}$ and an abundance of $0.11\, Z_\odot $, which are the values found from the {\it XMM-Newton} spectrum for one of the eROSITA-detected bona-fide period-bouncers \citep[see][]{stelzer2017}. In Table~\ref{fakeitmodel} we compare both conversion factors obtained using the simulation for a power law model ($CF_{\rm fake,powerlaw}$) and for an {\sc apec} model ($CF_{\rm fake,APEC}$) specifically for the eRASS:4 catalog (the values presented in \citealt{munoz2023} were obtained for a single-band eRASS:3 catalog), where it can be seen that these $CFs$ differ at most by $10$\,\%. This result allowed us to derive the $CF_{\rm eROSITA,APEC}$ from the $CF_{\rm eROSITA,powerlaw}$ used in the eRASS:4 catalog. 

The final values for the {\sc apec} fluxes for each eROSITA band (0.2-0.6 keV soft band, 0.6-2.3 keV medium band and 2.3-5.0 keV hard band) were calculated using the eRASS:4 catalog count rate and the {\sc apec} eROSITA conversion factor as $Flux_{ \rm APEC,i} = R_i/CF_{\rm eROSITA,APEC} $ where $i$=S,M,H and $R_i$ is the count rate in each band. The total {\sc apec} flux is calculated as the sum of the individual {\sc apec} fluxes from each band.  We combine the total {\sc apec} flux with the distance reported in the \cite{gentile2021} catalog to calculate the X-ray luminosity of each object. 


\begin{table}
    \caption{Rate-to-flux conversion factors for each eROSITA energy band in units of cts / erg * cm$^{2}$ * s. The conversion factors are calculated for simulated spectra obtained using an {\sc apec} model and a power law model with parameters detailed in Sect.~\ref{subsect:erositaCounterparts}). }        
    \label{fakeitmodel}     
    \centering                         
    \begin{tabular}{c c c c}       
    \hline              
    \noalign{\smallskip}
     & Soft band & Medium band & Hard band \\ 
     & 0.2-0.6 keV & 0.6-2.3 keV & 2.3-5.0 keV \\  
    \hline                       
    \noalign{\smallskip}
    $CF_{fake,powerlaw}$ & $3.61 \times 10^{11} $ & $4.61 \times 10^{11} $ & $3.33 \times 10^{10} $\\
    $CF_{fake,APEC}$ & $3.97 \times 10^{11} $ & $4.58 \times 10^{11} $ & $3.57 \times 10^{10} $\\
    \noalign{\smallskip}
    \hline                                  
    \end{tabular}
\end{table}


\subsection{ X-ray selection cuts}\label{subsect:erositaSelection}

\begin{figure}
    \centering
    \includegraphics[width=\columnwidth]{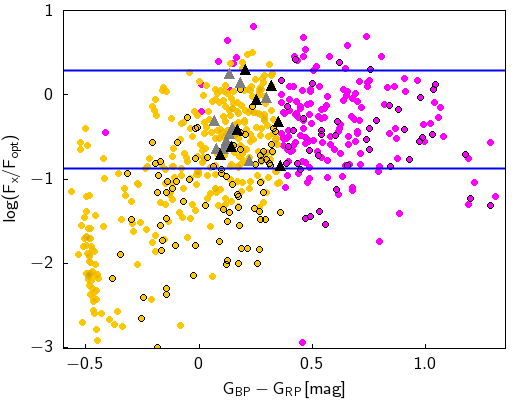}
    \caption{X-ray-to-optical flux ratio as a function of {\it Gaia} colors showing the position of the ``eROSITA WD subsample'' (circles) and confirmed period-bouncers (triangles) detected in eRASS:4. The bona-fide period-bouncers used to establish the selection cuts are highlighted in black. See text for the justification of the selection cut lines. The members of the ``eROSITA WD subsample'' that fulfill the X-ray luminosity and {\it Gaia} colors selection cuts are highlighted in yellow, the ones that do not are in pink. Black circles are placed around possible resolved binaries in the ``eROSITA WD subsample''.}
    \label{xrayfluxs4}
\end{figure}

\begin{figure}
    \centering
    \includegraphics[width=\columnwidth]{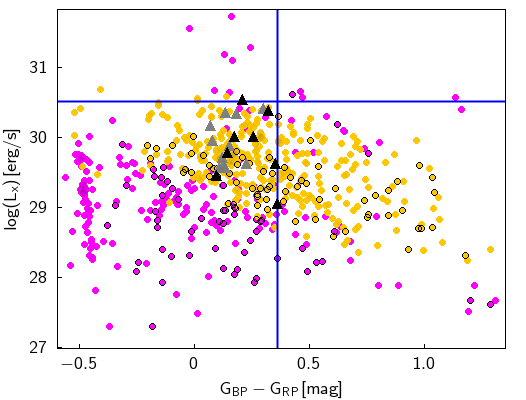}
    \caption{Bolometric X-ray luminosity as a function of {\it Gaia} colors showing the position of the ``eROSITA WD subsample'' (circles) and the bona-fide period-bouncers (triangles) detected in eRASS:4. The bona-fide period-bouncers used to establish the selection cuts are highlighted in black. See text for the justification of the selection cut lines. The members of the ``eROSITA WD subsample'' that fulfill the X-ray-to-optical flux ratio selection cut are highlighted in yellow, the ones that do not are in pink. Black circles are placed around possible resolved binaries in the ``eROSITA WD subsample''.}
    \label{luminosityS4}
\end{figure}

Following our earlier work \citep{munoz2024}, we use diagnostic diagrams that combine X-ray and optical data to establish selection cuts for period-bounce candidates based on the parameter space occupied by bona-fide period-bouncers. Because the number of bona-fide period-bouncers detected in each eROSITA catalog can vary, the X-ray selection cuts derived by \cite{munoz2024} for eRASS:3 are not necessarily applicable to eRASS:4. In Figs. \ref{xrayfluxs4} and \ref{luminosityS4} we have placed the ``eROSITA WD subsample'' in two diagnostic diagrams comprising eRASS data: X-ray-to-optical flux ratio ($\rm F_{\rm x}/F_{\rm opt}$) and X-ray luminosity (L$_x$) versus \textit{Gaia} color. The position of the ``eROSITA WD subsample'' is compared with the eROSITA-detected confirmed period-bouncers.



In Fig.~\ref{xrayfluxs4} more than half of the ``eROSITA WD subsample'' displays $-1 \leq \rm \log(F_x/F_{opt}) \leq 0$, typical of systems with low mass transfer rate including different CV types such as dwarf novae, polars and intermediate polars \citep{schwope2024}. This parameter on its own does not identify potential period-bouncers, but it serves to discard systems with high optical flux indicative of either high mass transfer rate CVs or hot single white dwarfs. The accumulation of members of the ``eROSITA WD subsample'' towards the bottom left corner of Fig.~\ref{xrayfluxs4} all have WD temperatures reported by \cite{gentile2021} hotter than 25000K, more than twice the expected value for period-bouncers, with $34$ of them already being securely identified as single WDs \citep{friedrich2025}.

In Fig.~\ref{luminosityS4} almost all of the ``eROSITA WD subsample'' displays L$_x \leq 10^{31}$ erg/s, characteristic of period-bouncers and low-mass transfer systems. The ``eROSITA WD subsample'' shows a broad distribution in \textit{Gaia} color which, excluding the distinct group around $G_{BP}-G_{RP}\cong$0.5 discussed above, represent systems dominated by cool WDs with effective temperatures between 7000K and 15000K.

X-ray plus \textit{Gaia} color selection cuts (see blue lines in Figs. \ref{xrayfluxs4} and \ref{luminosityS4}) are defined on the basis of the position of eROSITA-detected bona-fide period-bouncers in Figs. \ref{xrayfluxs4} and \ref{luminosityS4} as discussed above, and are shown as blue lines resulting in upper and lower limits for the X-ray-to-optical flux ratio ($-0.87 \leq \rm \log(F_x/F_{opt}) \leq 0.28$), and in upper limits for the X-ray luminosity ($\log$(L$_{\rm x})\leq$ 30.52 [erg/s]) and {\it Gaia} color ($G_{BP}-G_{RP}\leq$ 0.36). These three criteria are fulfilled by $213$ members of the ``eROSITA WD subsample'', which are henceforth considered period-bounce candidates. 



\begin{figure}
    \centering
    \includegraphics[width=\columnwidth]{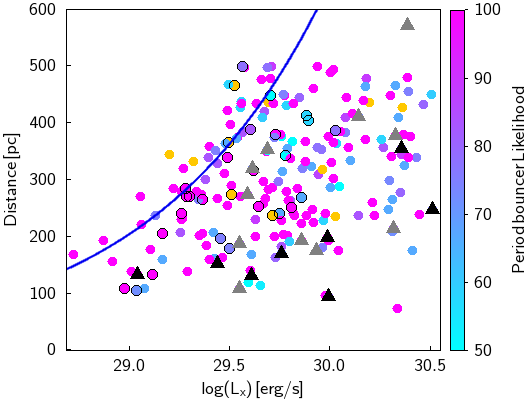}
    \caption{Distance from \cite{bailer2021} versus X-ray luminosity showing the position of the $213$ period-bounce candidates (in color scale) and the eROSITA-detected confirmed period-bouncers (in triangles, bona-fide period-bouncers are highlighted in black). The average eRASS:4 sensitivity limit is marked by the blue line (see text in Sect.~\ref{sect:PBcandidates}). Black borders are used to show possible resolved binaries in the ``eROSITA WD subsample''.}
    \label{distaceLogls4}
\end{figure}

\section{High-likelihood period-bounce candidates}\label{sect:PBcandidates}

We show in Fig.~\ref{distaceLogls4} the relationship between X-ray luminosity and distance for the period-bounce candidates identified in Sect.~\ref{subsect:erositaSelection}. Using a conservative eRASS:4 detection likelihood of {\sc det\_ml} $>10$, which corresponds to the value for 2QZ J14283+0031 (the lowest amongst eROSITA-detected confirmed period-bouncers), $95$\,\% of the sources in the full eRASS:4 catalog have a flux higher than $2 \times 10^{-14}\,{\rm erg/cm^2/s}$, which is used to calculate the average eRASS:4 sensitivity limit shown in Fig.~\ref{distaceLogls4} as a blue line. Our new period-bounce candidates tend to have, at a given distance, even lower X-ray luminosity when compared to eROSITA-detected confirmed period-bouncers. X-ray faint sources are of special interest in our search for period-bouncers considering that they might have remained undetected in other multiwavelength studies for the missing members of this population.

To test whether the $213$ period-bounce candidates, especially the ones around the eRASS:4 sensitivity limit, constitute the missing population of period-bouncers we apply the reduced scorecard introduced in Sect.~\ref{subsect:reducedScorecard} to this population. The calculated final scores are presented as a color scale in Fig.~\ref{distaceLogls4}. Period-bounce candidates that have reported values only for two or less of the parameters from the reduced scorecard do not have enough information for us to judge accurately their status as a period-bouncer an therefore are not assigned a final score (yellow symbols in Fig.~\ref{distaceLogls4}).

The eROSITA-detected bona-fide period-bouncer with the lowest final score calculated using the reduced scorecard is V406\,Vir (also the lowest scoring bona-fide period-bouncer in \citealt{munoz2024} and \citealt{munoz2024b}) with a reduced final score of 72$\%$. This means that a category of high-likelihood period-bouncers can be defined as systems that have final scores higher than this value. This applies to $161$ systems which we consider high-likelihood period-bounce candidates. It is relevant to mention that a high fraction of these high-likelihood period-bounce candidates, $108$ systems, have a final score of 100$\%$ according to the reduced scorecard (systems shown in pink in Fig.~\ref{distaceLogls4}). 



\subsection{Classification}\label{subsect:PBcandidatesClassification}

A cross-match with SIMBAD gives us the classification of the $161$ high-likelihood period-bounce candidates according to the previous literature. The J2000 coordinates (given in epoch 2016.0) available from the \textit{Gaia} DR3 catalog were used for a cross-match with SIMBAD\footnote{SIMBAD stands for Set of Identifications, Measurements and Bibliography for Astronomical Data, available at \url{http://simbad.cds.unistra.fr/simbad.}}. We verified that the object retrieved from the match had the same \textit{Gaia} DR3 {\sc source\_ID} as the one reported in the \cite{gentile2021} catalog. We give here more information on the SIMBAD classification:

{\it Period-bouncers:} 15 of the high-likelihood period-bounce candidates are already known systems classified as period-bouncers, including 6 eROSITA-detected bona-fide period-bouncers and 9 eROSITA-confirmed period-bouncers. The successful retrieval of these systems from the \cite{gentile2021} WD catalog conclusively proves the usefulness of our selection tools (period-bouncer scorecard and X-ray selection cuts) to identify period-bouncers hiding in WD catalogs. 

{\it Cataclysmic variables:} 5 of the high-likelihood period-bounce candidates had already been classified as CVs. This includes 1 known polar (CP Tuc), 1 known WZ Sge (ASSASN -17el), 1 known nova (ASASSN -18fk) and 2 without further classification (PM J11384+0619 ; Gaia 18ctg). ASASSN -17el is the only system that is included in the catalog of CVs around the period-bounce by \cite{munoz2024} and was initially not confirmed as a period-bouncer due to its X-ray luminosity in eRASS:3 being slightly higher than the X-ray selection cut used, however considering that it falls within the selection cut for eRASS:4 we can now confirm it as a new period-bouncer. The other 4 systems were not included in the catalog of CVs around the period-bounce by \cite{munoz2024} or its addition \citep{munoz2024b} as they had not been previously suggested as period-bouncers and did not have an orbital period or donor mass placing them in the period-bounce area. We hereby confirm all 4 of these systems (CP Tuc, ASASSN -18fk, PM J11384+0619, and Gaia 18ctg) as new period-bouncers considering that they have an IR excess and multiwavelength colors (GALEX, SDSS and IR) additional to their eROSITA and {\it Gaia} data that support their classification as a period-bouncer, as well as photometrically detected very late-type donors (see Appendix~\ref{appen:SED}).

{\it AM CVn binaries:} 2 of the high-likelihood period-bounce candidates are already known AM CVn binaries (V396 Hya ; GP Com). This is a distinct sub-class of CVs that underwent common envelope evolution twice, rather than the single common envelope evolution that the rest of the CVs undergo \citep{nelemans2001}. AM CVn binaries are known for containing a WD accreting from a hydrogen-deficient donor star that is fully or partially degenerate \citep{deloye2005}. They exhibit orbital periods from 5\,min to 68\,min (\citealt{levitan2015}, \citealt{ramsay2018}, \citealt{green2020}). Similar to period-bouncers, AM CVn binaries are heavily underrepresented in observational surveys for CVs \citep{rodriguez2025}. From this result, we can anticipate that a small number of our high-likelihood period-bounce candidates will turn out to be AM CVn binaries, meaning that our search will also contribute to the characterization of another underrepresented population.

{\it RR Lyrae Variable Candidate}: 1 of the high-likelihood period-bounce candidates has been tentatively identified as a RR Lyr variable candidate based on its {\it Gaia} data (Gaia DR3 2930889294867085440; \citealt{clementini2019}). Because there is no additional information of this system apart from eROSITA and {\it Gaia} data, we can not securely classify it as a CV or a period-bouncer. Follow-up of this target is required to proceed with its confirmation as a period-bouncer.

{\it WDs and WD candidates:} The remaining $138$ high-likelihood period-bounce candidates are classified as either a WD or a WD candidate with $\sim\,80\%$ of them only appearing in the literature as part of large WD catalogs (\citealt{jimenez2018}, \citealt{gentile2019}, \citealt{gentile2021}, \citealt{torres2023}, \citealt{vincent2024}). One confirmed period-bouncer (eRASS J05472+1326; \citealt{rodriguez2025}) is incorrectly classified as a WD in SIMBAD, such that we do not need to consider it further in our effort to confirm new period-bouncers. These candidates constitute the focus of the following sections.

\subsection{Confirmation of high-likelihood period-bounce candidates} \label{subsect:PBcandidatesConfirmation}

We are left with $137$ high-likelihood period-bounce candidates that are classified as WDs or WD candidates. The confirmation of these candidates is imperative for the discovery of the missing population of period-bouncers. Here we describe what are the required steps for a secure period-bouncer classification, followed by the presentation of a newly identified period-bouncer using this approach.

\subsubsection{Requirements for confirmation as a period-bouncer} \label{subsubsect:3stepPBconfirmation}

In order to securely confirm the $137$ high-likelihood candidates classified as WDs or WD candidates as period-bouncers we must fulfill three requirements: confirmation of the system as a CV, determination of an orbital period, and detection of a late-type donor.

{\it Confirmation as a CV:} In order to confirm a system as a period-bouncer we must first make sure that it is a CV, meaning that it is an interacting/accreting binary rather than a single WD. The most straightforward method to do so is through the detection of Balmer emission lines in the optical spectra of potential CV candidates \citep{warner2003}, which is a clear indicator of accretion in the system, and conclusively differentiates them from single WDs which only present Balmer absorption lines \citep{wesemael1993}. The SDSS-V Project is pioneering all-sky panoptic\footnote{Panoptic: presenting a comprehensive or encompassing view of the whole} spectroscopy by providing the first optical plus infrared survey for millions of sources spread across the entire sky \citep{kollmeier17, kollmeier25}.  One of the SDSS-V subprojects aims to spectroscopically observe up to 60000 WDs. The survey began its run in November 2020, initially using the 2.5 m telescope at the Apache Point Observatory \citep{gunn06} and has now expanded to 2.5 m telescope at Las Campanas Observatory \citep{bv73}, and observes targets using the Baryon Oscillation Spectroscopic Survey spectrograph \citep[BOSS;][]{smee13} and the Apache Point Observatory Galactic Evolution Experiment spectrographs \citep[APOGEE; ][]{wilson19}. Most of the WDs are observed as part of the multi-epoch Milky Way Mapper program \citep[MWM; ][]{kollmeier25}.  Each SDSS-V BOSS sub-exposure takes 900s, has a median resolution of $R\sim1800$, and covers the wavelength range from 3600 \AA to 10000 \AA. These SDSS follow-up spectra will help us confirm or deny the classification of our high-likelihood candidates as CVs, as well as help us establish if the WD is magnetic and parameters like WD temperature, mass and radius (see Appendix~\ref{appen:WDtemperature}). Similar searches for CVs and period-bouncers, however only focused on optical wavelenghts, have been conducted using SDSS (\citealt{inight2023b}, \citealt{inight2023a}) which has lead to the discovery of a large number of new systems.

{\it Orbital period determination:} The period-bounce candidate has to present an orbital modulation associated with a close interacting binary, that places the system around the period minimum of CVs ($\sim$80 min; \citealt{gansicke2009}). In \cite{hernandez2025} we demonstrated that the two-minute cadence light curves from the {\it Transiting Exoplanet Survey Satellite} (TESS; \citealt{TESS}) are well-suited for efficiently detecting orbital periods of CVs. We validated the effectiveness of our methodology and introduced a probabilistic framework to assess the reliability of period detections with TESS using the parameter SNR$_{\text{PSD}}$, the signal-to-noise ratio (SNR) of the frequency peak in the Power Spectral Density (PSD) corresponding to the orbital period. Our approach, which involves four different techniques to measure orbital periods, can now be applied to any other sample of CV candidates that have TESS two-minute cadence light curves available. 

{\it Detection of late-type donor:} The defining characteristic of period-bouncers is the presence of a very late-type degenerate donor, which ultimately is what distinguishes them from the rest of CVs. The detection of a very late-type donor can be achieved spectroscopically or photometrically. A spectroscopic detection results in the direct detection of the donor through observation of specific absorption lines, with a the secure determination of the spectral type of the donor. On the other hand, a photometric detection results in an approximation to the spectral type of the donor based on the IR excess presented by the candidate in its spectral energy distribution (SED). As detailed by \cite{munoz2024}, a spectroscopic detection of a very late-type donor is the ultimate confirmation of a CV as a period-bouncer, with a photometric detection only used as definitive confirmation when other evidence supports the classification of the system as a period-bouncer. 

\subsubsection{GALEX J125751.4-283015: A new period-bouncer}\label{subsubsect:galex1257}

\begin{figure*}
    \centering
    \includegraphics[width=0.56\textwidth]{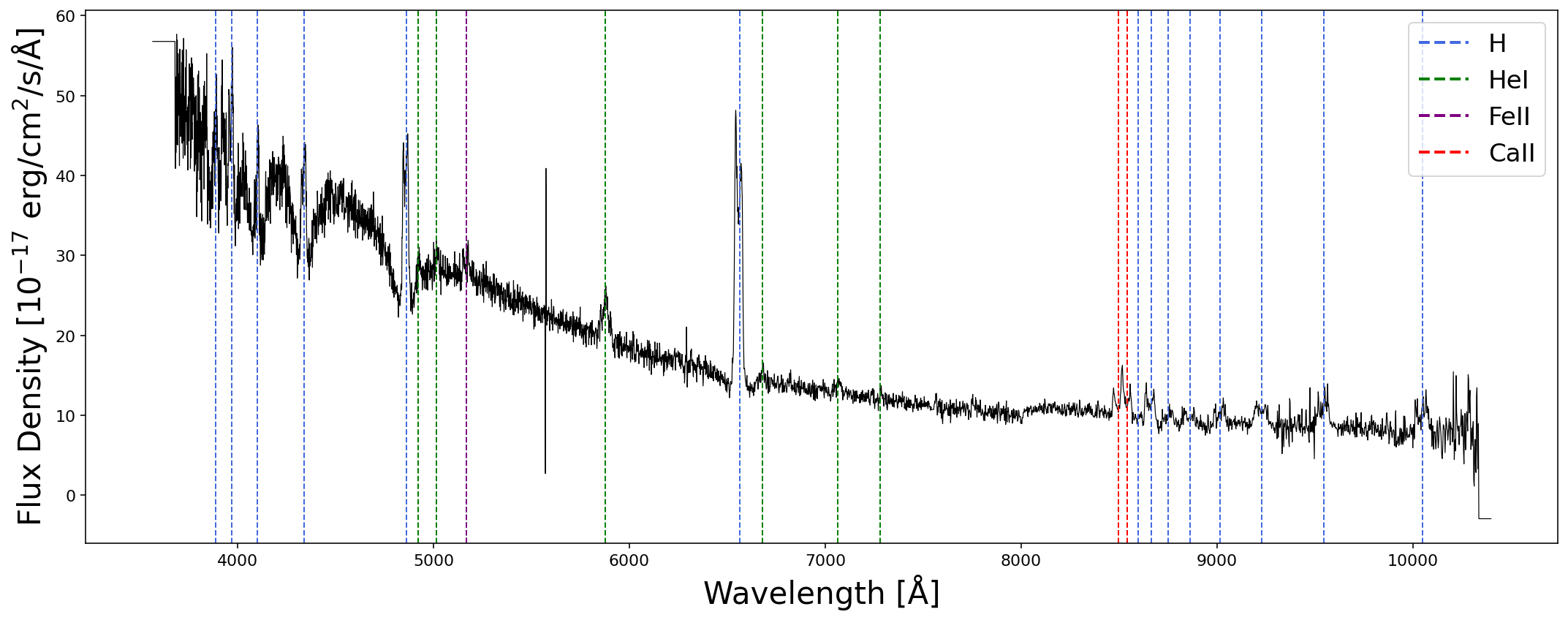}\hfill
    \includegraphics[width=0.43\textwidth,trim=0.1cm 0.5cm 1.69cm 1.2cm,clip]{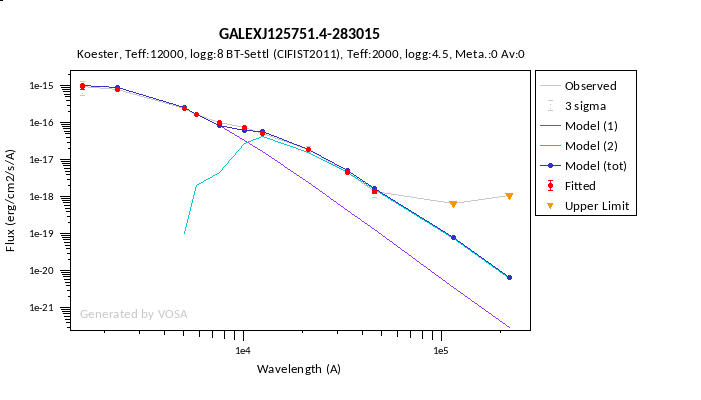}
    \includegraphics[width=0.48\textwidth,trim=0.2cm 0.2cm 0.2cm 0.2cm,clip]{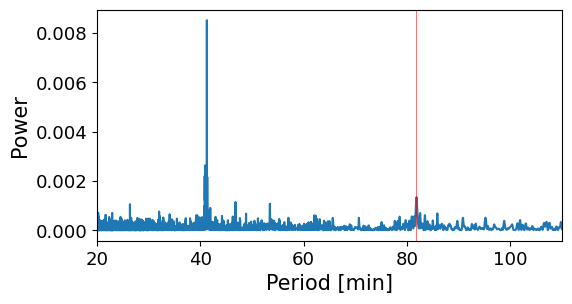}\hfill
    \includegraphics[width=0.51\textwidth,trim=0.3cm 0.4cm 0.7cm 0.3cm,clip]{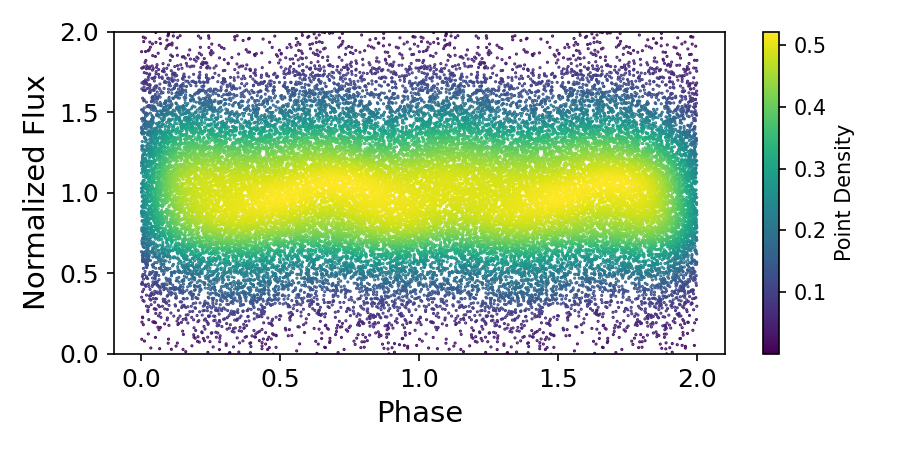}
    \includegraphics[width=\textwidth,trim=6.2cm 0cm 5.6cm 1.4cm,clip]{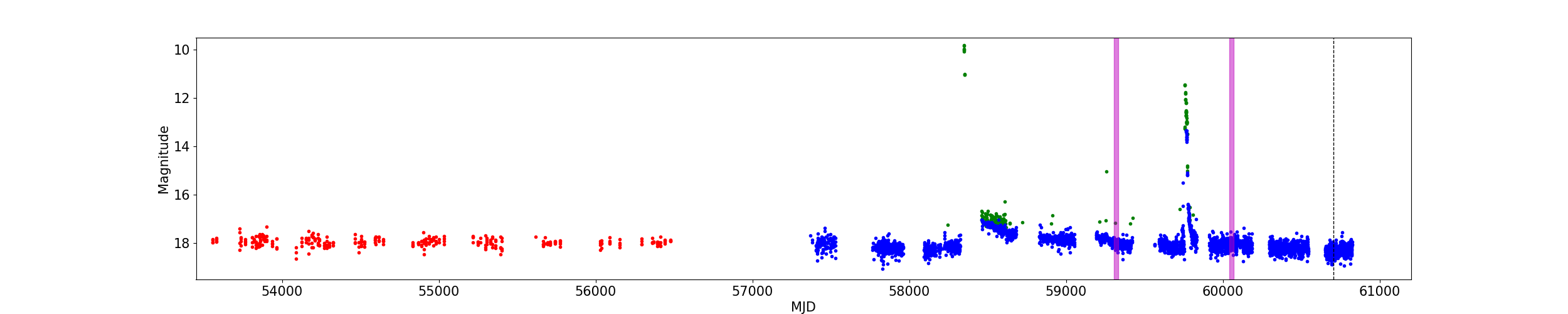}
    \caption{Newly confirmed period-bouncer GALEX J125751.4-283015. \textit{Upper left panel:} SDSS spectrum showing clear Balmer emission lines. \textit{Upper right panel:} Spectral energy distribution fitted using a hydrogen WD model (T$_{\rm eff}$ = 12000\,K and $\log(g)$=8.0) and a late-type star model (T$_{\rm eff}$ = 2000\,K and $\log(g)$=4.5). \textit{Middle left panel:} Peaks obtained for orbital period and half orbital period. \textit{Middle right panel:} Phase folded lightcurve (sector\,64) for a period of 82.5258\,min. \textit{Lower panel:} Combined light curves from CRTS (red), ATLAS (blue) and ASAS-SN (green). Vertical purple stripes show the epochs of the TESS Sector\,37 and Sector\,64 light curves, while the vertical dashed black line shows the the epoch of the SDSS spectrum.}
    \label{GALEX1257}
\end{figure*}

In a preliminary, not yet systematic study following the three-fold strategy outlined in Sect.~\ref{subsubsect:3stepPBconfirmation}, we identified GALEX\,J125751.4-283015 as a new period-bouncer based on its SED, optical light curve and spectrum.

GALEX\,J125751.4-283015 is classified according to SIMBAD as a WD candidate. It has been suggested as the probable progenitor of ASASSN -18su \citep{barba2018}, a bright transient classified as a non-magnetic CV \citep{le2015} thanks to an optical spectrum obtained with the NTT/EFOSC2 (New Technology Telescope/ESO Faint Object Spectrograph and Camera v.2). 

Two optical spectra of GALEX\,J125751.4-283015 were obtained in the framework of the SDSS-V follow-up program outlined in Sect.~\ref{subsubsect:3stepPBconfirmation}. The SDSS spectra further supports the classification as a CV (see Fig.~\ref{GALEX1257} upper left panel) showing clear Balmer emission lines characteristic of accreting binaries. As is typical of quiescent dwarf novae, the strong Balmer emission lines are accompanied by emission of Paschen lines, He\,I, Ca\,II and Fe\,II (\citealt{mason2005}, \citealt{southworth2009}). Specially the Balmer emission lines exhibit a clear double-peaked profile, arising from the rotating Keplerian disk that generates the Doppler shift (\citealt{smak1969,smak1982}, \citealt{horne1986}).
Each of the two SDSS spectra was combined with the GALEX photometry in order to derive the WD parameters (see Appendix~\ref{appen:WDtemperature}) resulting in an effective temperature between 12000K to 12500K and a $\log(g)$ between 8.0 and 8.5, which are typical ranges for CVs near the period minimum (\citealt{pala2020,pala2022}). We also obtain a WD radius between 8.01$\times 10^8$\,cm and 8.93$\times 10^8$\,cm that, when using the mass-radius relation by \cite{nauenberg}, results in a WD mass between 0.67M$_{\odot}$ and 0.57M$_{\odot}$ respectively.

GALEX\,J125751.4-283015 has the identifier TIC\,999085360 in the TESS Input Catalog (\citealt{TIC}), with two TESS two-minute cadence light curves, from sectors\,37 and 64, available from the Barbara A. Mikulski Archive for Space Telescopes (MAST)\footnote{\citealt{mast_portal}}. These light curves were used to derive an orbital period of $82.17\pm0.63$.

In the Lomb-Scargle periodogram of the TESS two-minute cadence light curve from sector $64$ (middle panels in Fig.~\ref{GALEX1257}), two significant frequency peaks can be identified at 41.26\,min and 81.83\,min, while in the TESS two-minute cadence light curve from sector $37$, only a signal at 41.26\,min is detected. The signal at 41.26\,min in sector\,$64$ shows the highest significance with ${\rm SNR}_{\text{PSD}} = 0.011$, followed by the detections at 81.83\,min in sector\,$64$ and at 41.26\,min in sector $37$ which have ${\rm SNR}_{\text{PSD}} = 0.0078$ and ${\rm SNR}_{\text{PSD}} = 0.0077$, respectively. All these values exceed the detection threshold of $({\rm SNR}_{\text{PSD}})_{\rm min} = 0.004$ established by \cite{hernandez2025}, supporting the reliability of these detections. 

We interpret the 41.26\,min signal as the second harmonic, and associate the 81.83\,min signal to the orbital period. Double-humped modulations in non-magnetic CVs can arise from the ellipsoidal modulation of the secondary (see \citealt{Wilson2006} and \citealt{Bochkarev1979}), from partial eclipses between the accretion disk and the secondary in high inclination systems (see \citealt{Cynthia_1999}) or from variations in the projected area of an elongated hotspot which remains visible throughout the whole orbital cycle (expected when the disk is optically thin and the hotspot optically thick), producing two brightness maxima per orbit (see e.g. \citealt{skidmore_2000}). In addition, the 2:1 resonance radius of the disk (see \citealt{Kunze_2005}) can drive two spiral density waves in its outer ring, also producing a double-humped shape in the observed light curve (see \citealt{Aviles_2010}, \citealt{zharikov2013} and \citealt{amantayeva2021}). This phenomenon is expected to happen in short-period CVs with low mass ratios ($q \lesssim 0.1$), where the enlarged Roche lobe of the primary allows the accretion disk to extend out to the 2:1 resonance radius.



The long-term lightcurve of GALEX J125751.4-283015 was constructed using optical photometry from the Catalina Real-time Transient Survey (CRTS; \citealt{drake2009}), Asteroid Terrestrial-impact Last Alert System (ATLAS; \citealt{tonry2018}) and All-Sky Automated Survey for Supernovae (ASAS-SN; \citealt{shappee2014}, \citealt{kochanek2017}) surveys (see Fig.~\ref{GALEX1257} lower panel). The lightcurve shows that GALEX J125751.4-283015 remains the majority of the time in a quiescent state (at $\sim$18mag) with only two records of enhanced activity from both ATLAS and ASAS-SN, a superoutburst in August 2018 lasting around 9 months with a maximum magnitude of $\sim$9.8 and an outburst in May 2022 lasting around 3 months with a maximum magnitude of $\sim$11.4. This low number of outbursts recorded in around 20 years of data is indicative of low mass accretion rate systems like period-bouncers. The TESS two-minute cadence light curves as well as the SDSS spectrum are taken during periods of quiescence, and can therefore be used to derive reliable parameters of the system.

Considering that there is no available spectra of GALEX J125751.4-283015 covering the infrared wavelengths, our approach is to use the SED (constructed as detailed in \citealt{munoz2024}) to obtain an approximate spectral type for the donor. Photometry from GALEX, \textit{Gaia}, VHS and AllWISE are used to construct the SED of GALEX J125751.4-283015 (see Fig.~\ref{GALEX1257} upper right panel). To derive the parameters of the individual components from the SED we used the Virtual Observatory SED Analyzer (VOSA, \citealt{bayo2008}). We initially fitted a WD model \citep{koester2010} to the SED giving as input a range of $\pm$1000\,K around the WD temperature of 11750\,K, the lowest value we derived from the SDSS spectrum and GALEX photometry (see Appendix~\ref{appen:WDtemperature}). 
The final WD parameters from the best fit are T$_{\rm eff}$=12000\,K and $\log(g)$=8.0, with VOSA suggesting an excess starting in the WISE1 band (consistent with findings for known period-bouncers; \citealt{munoz2024}). The model binary fit tool from VOSA allows fitting of the SED simultaneously with two models, in our case a WD model \citep{koester2010} and a late-type star model (BT-Settl; \citealt{allard2003}). We fix the WD parameters on the previously derived values. The best fit binary model results in a secondary component with T$_{\rm eff}$=2000\,K and $\log(g)$=4.5. In the case that the entire contribution of the secondary component can be attributed to the donor star, it would correspond to an spectral type L0 according to the standard evolutionary track for CVs \citep{knigge2011}. However, considering that it is very likely that part of this secondary contribution is coming from the accretion disk (\citealt{pala2018}, \citealt{neustroev2022}), we can characterize the spectral type of the donor as L0 or later.

To summarize, according to our multiwavelength analysis GALEX\,J125751.4-283015 is a confirmed CV with an orbital period of $82.17\pm0.63$\,min and a donor with a spectral type L0 or later, meaning that this high-likelihood candidate is most likely a period-bouncer. A future spectroscopic detection and characterization of the donor will allow GALEX J125751.4-283015 to become part of the sample of bona-fide period-bouncers.





\section{Conclusions}\label{sect:conclusions}

With this work we have conclusively proven that there are previously unknown WD-dominated CVs hiding amongst WD catalogs which can be identified through the use of X-rays, and constitute a sample of high-likelihood period-bounce candidates. The application of similar search strategies have already produced the recent discovery of two new period-bouncers (WD J18204-0422 and WD J19070+2052; \citealt{cunningham2025}) which were identified from a search for soft X-ray emission from likely single WDs in the \textit{XMM-Newton} source catalogue. 

We explored the possible X-ray emission of \textit{Gaia}-identified WDs and found $601$ of them that appear to have a reliable eROSITA counterpart, constituting the ``eROSITA WD subsample''. This subsample includes several single WD candidates with effective temperatures larger than 25000K, twice as much as the expectations for period-bouncers. \cite{friedrich2025} used the eROSITA X-ray emission of systems in the \cite{gentile2021} WD catalog, to identify 264 single WD candidates in the western Galactic half sky. Our parallel confirmation campaigns should help establish if the single WD candidates are indeed isolated objects or CVs. Aditionally, \textit{Gaia} astrometry of members of the ``eROSITA WD subsample'' reveals that $122$ most likely are part of a resolved binary. The X-ray emission in this case could be associated with the wide companion (not the WD, meaning it is most likely a single cool WD), or could be indeed originating from an unresolved interacting binary in a triple system configuration. \cite{shariat2025} found $\sim$50 CVs in such configurations, including two confirmed period-bouncers (V1108 Her and 1RXS J10142+0638), making it likely that we might find such systems in our search for period-bouncers. Future detailed studies of these potential binaries in our ``eROSITA WD subsample'' will reveal if they are a triple system with an inner CV or a wide WD binary system. 

The selection cuts and a reduced version of the period-bouncer scorecard by \cite{munoz2024} are applied to the ``eROSITA WD subsample'' to identify $161$ high-likelihood period-bounce candidates. Five high-likelihood period-bounce candidates that had already been classified as CVs in the literature (CP Tuc, ASSASN -17el, ASSASN -18fk, PM J11384+0619, and Gaia 18ctg) are now confirmed as new period-bouncers, with photometrically detected donors ranging from spectral types L9.3 to T. Future detailed spectroscopic studies will help confirm the spectral type of the donors, and shed more light on their status as period-bouncers. Follow-up of Gaia DR3 2930889294867085440, a RR Lyr variable candidate, is required to proceed with its confirmation as a period-bouncer.

The ongoing effort to confirm high-likelihood period-bounce candidates that have only been identified as WDs in the literature has produced its first successful case, with the re-classification of GALEX\,J125751.4-283015 as a period-bouncer. This proves the capabilities of our period-bounce scorecard and selection cuts to identify period-bouncers hiding in WD catalogs, suggesting that there are several additional systems waiting to be discovered that could represent the missing population of period-bouncers.

The addition of six newly confirmed period-bouncers (five previously known CVs and one WD candidate) represents an increase in the population of some$\,20 \%$, bringing the number of this elusive class of CVs to $39$ systems. Within 500pc period-bouncers now make-up 17$\%$ of eROSITA-detected CVs, comparable to the 7-25$\%$ fraction estimated for a 150pc volume limited sample of CVs (\citealt{pala2020}, \citealt{rodriguez2025}). Considering that the known population of period-bounce CVs still remains well below the numbers expected by theoretical models, we foresee that further exploitation of eROSITA data will boost the population number even to the predicted levels, especially considering that the confirmation of high-likelihood period-bounce candidates around the eROSITA sensitivity limit is on its way.

The confirmation of the remaining $136$ high-likelihood period-bouncers with a WD classification from SIMBAD, which is presently ongoing, is the most likely path for finding the missing population of period-bouncers. This could result in the validation of present-day evolution models for CVs without requiring a revision of the predicted numbers of systems.


\begin{acknowledgements}
Daniela Muñoz-Giraldo was supported by Deutsche Forschungsgemeinschaft (DFG) under grant number STE 1068/6-1. This work is based on data from eROSITA, the soft X-ray instrument aboard SRG, a joint Russian-German science mission supported by the Russian Space Agency (Roskosmos), in the interests of the Russian Academy of Sciences represented by its Space Research Institute (IKI), and the Deutsches Zentrum f\"{u}r Luft- und Raumfahrt (DLR). The SRG spacecraft was built by Lavochkin Association (NPOL) and its subcontractors, and is operated by NPOL with support from the Max Planck Institute for Extraterrestrial Physics (MPE). The development and construction of the eROSITA X-ray instrument was led by MPE, with contributions from the Dr. Karl Remeis Observatory Bamberg \& ECAP (FAU Erlangen-Nuernberg), the University of Hamburg Observatory, the Leibniz Institute for Astrophysics Potsdam (AIP), and the Institute for Astronomy and Astrophysics of the University of T\"{u}bingen, with the support of DLR and the Max Planck Society. The Argelander Institute for Astronomy of the University of Bonn and the Ludwig Maximilians Universit\"{a}t Munich also participated in the science preparation for eROSITA. The eROSITA data shown here were processed using the eSASS/NRTA software system developed by the German eROSITA consortium. This work has made use of data from the European Space Agency (ESA) mission {\it Gaia} (\url{https://www.cosmos.esa.int/gaia}), processed by the {\it Gaia} Data Processing and Analysis Consortium (DPAC, \url{https://www.cosmos.esa.int/web/gaia/dpac/consortium}). Funding for the DPAC has been provided by national institutions, in particular the institutions participating in the {\it Gaia} Multilateral Agreement. This publication makes use of VOSA, developed under the Spanish Virtual Observatory (\url{https://svo.cab.inta-csic.es}) project funded by MCIN/AEI/10.13039/501100011033/ through grant PID2020-112949GB-I00. VOSA has been partially updated by using funding from the European Union's Horizon 2020 Research and Innovation Programme, under Grant Agreement n$^\circ$ 776403 (EXOPLANETS-A).

Funding for the Sloan Digital Sky Survey V has been provided by the Alfred P. Sloan Foundation, the Heising-Simons Foundation, the National Science Foundation, and the Participating Institutions. SDSS acknowledges support and resources from the Center for High-Performance Computing at the University of Utah. SDSS telescopes are located at Apache Point Observatory, funded by the Astrophysical Research Consortium and operated by New Mexico State University, and at Las Campanas Observatory, operated by the Carnegie Institution for Science. The SDSS web site is \url{www.sdss.org}. SDSS is managed by the Astrophysical Research Consortium for the Participating Institutions of the SDSS Collaboration, including the Carnegie Institution for Science, Chilean National Time Allocation Committee (CNTAC) ratified researchers, Caltech, the Gotham Participation Group, Harvard University, Heidelberg University, The Flatiron Institute, The Johns Hopkins University, L'Ecole polytechnique f\'{e}d\'{e}rale de Lausanne (EPFL), Leibniz-Institut f\"{u}r Astrophysik Potsdam (AIP), Max-Planck-Institut f\"ur Astronomie (MPIA Heidelberg), Max-Planck-Institut f\"{u}r Extraterrestrische Physik (MPE), Nanjing University, National Astronomical Observatories of China (NAOC), New Mexico State University, The Ohio State University, Pennsylvania State University, Smithsonian Astrophysical Observatory, Space Telescope Science Institute (STScI), the Stellar Astrophysics Participation Group, Universidad Nacional Aut\'{o}noma de M\'{e}xico, University of Arizona, University of Colorado Boulder, University of Illinois at Urbana-Champaign, University of Toronto, University of Utah, University of Virginia, Yale University, and Yunnan University.

\end{acknowledgements}

%
%

\bibliography{references.bib}
\bibliographystyle{aa} 

\begin{appendix}
\section{Photometric detection of a late-type donor}\label{appen:SED}

Similar to the methodology described in \cite{munoz2024}, we used the photometry available for the newly confirmed period-bouncers in order to prove the presence of a late-type donor in the system. The VOSA fit using a WD model \citep{koester2010} results in the WD effective temperatures shown in Table~\ref{TeffNewPBs}, which all agree within 18$\%$ of the published value available for two of them. The obtained WD temperature is then used for the binary fit that characterizes the excess emission over the contribution of the WD. Considering that the excess emission can be coming fully or partially from the donor this corresponds to a L9.3-type or later donor for ASASSN -17el, and a T-type donor for the other new period-bouncers \citep{knigge2011}. 


\begin{table}
    \begin{threeparttable}
    \centering
    \caption{WD temperature obtained for new period-bouncers using VOSA compared to values found in the literature.}
    \label{TeffNewPBs}
    \begin{tabular}{lcc} 
    \hline
    \noalign{\smallskip}
     & VOSA & Literature  \\ 
     & [K] & [K]  \\  
    \hline                       
    \noalign{\smallskip}
    CP Tuc & 11750  & 10000 $\pm$ 1500 (1) \\
    ASSASN -17el & 12000 &   \\
    ASSASN -18fk & 12000 & 11000 $\pm$ 500 (2) \\
    PM J11384+0619 & 11500 &  \\
    Gaia 18ctg & 12250 &  \\
    \hline
    \end{tabular}
    \begin{tablenotes}
        \small
        \item {\it References.} (1) \cite{beuermann2007}, (2) \cite{wang2020}.
    \end{tablenotes}
  \end{threeparttable}
\end{table}


\begin{figure}
    \centering
    \includegraphics[width=\columnwidth,trim=0.1cm 0.5cm 1.1cm 1.2cm,clip]{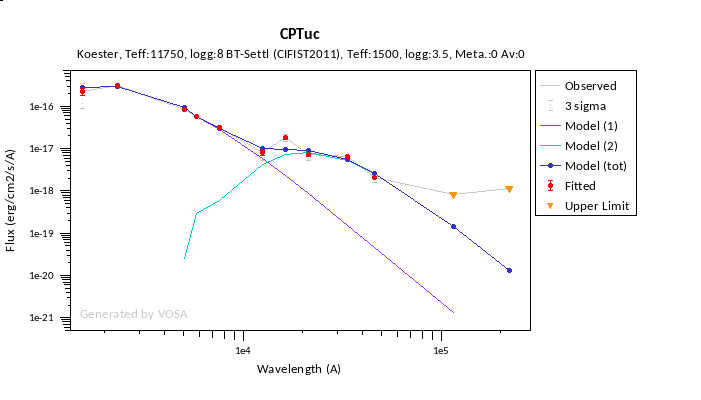}
    \caption{Binary fit to the SED of CP Tuc. The temperature for the secondary component is indicative of a donor with a spectral type T.}
    \label{SEDcptuc}
\end{figure}
\begin{figure}
    \centering
    \includegraphics[width=\columnwidth,trim=0.1cm 0.5cm 1.1cm 1.2cm,clip]{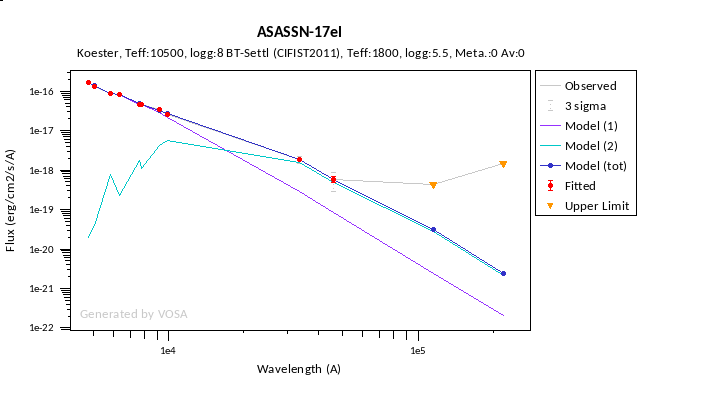}
    \caption{Binary fit to the SED of ASASSN -17el. The temperature for the secondary component is indicative of a donor with a spectral type L9.3.}
    \label{SEDasassn-17el}
\end{figure}
\begin{figure}
    \centering
    \includegraphics[width=\columnwidth,trim=0.1cm 0.5cm 1.1cm 1.2cm,clip]{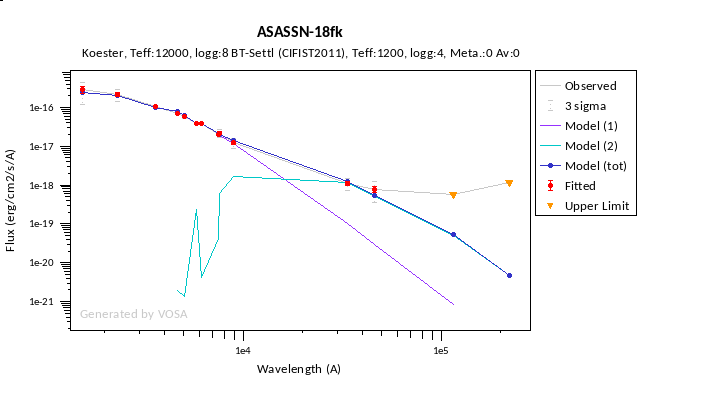}
    \caption{Binary fit to the SED of ASSASN -18fk. The temperature for the secondary component is indicative of a donor with a spectral type T.}
    \label{SEDasassn-18fk}
\end{figure}
\begin{figure}
    \centering
    \includegraphics[width=\columnwidth,trim=0.1cm 0.5cm 1.1cm 1.2cm,clip]{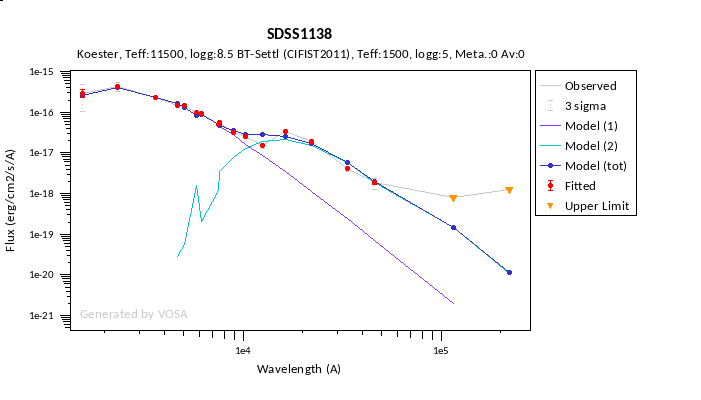}
    \caption{Binary fit to the SED of PM J11384+0619. The temperature for the secondary component is indicative of a donor with a spectral type T.}
    \label{SEDpm1138}
\end{figure}
\begin{figure}
    \centering
    \includegraphics[width=\columnwidth,trim=0.1cm 0.5cm 1.1cm 1.2cm,clip]{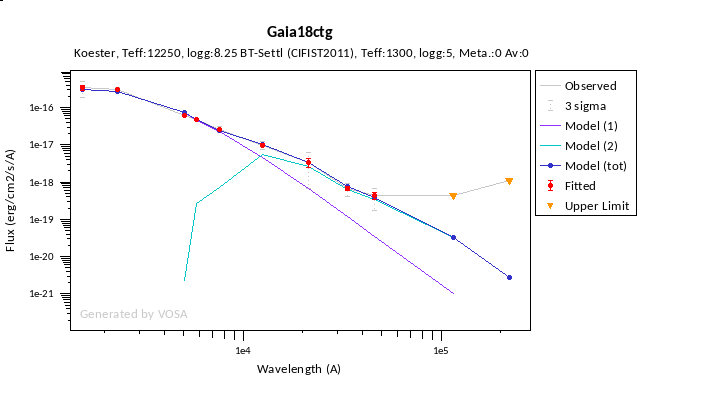}
    \caption{Binary fit to the SED of Gaia 18ctg. The temperature for the secondary component is indicative of a donor with a spectral type T.}
    \label{SEDgaia18ctg}
\end{figure}

\section{WD parameters from SDSS spectra and GALEX photometry}\label{appen:WDtemperature}

\begin{figure*}
    \centering
    \includegraphics[width=\textwidth,trim=0.4cm 1.5cm 0.3cm 0.2cm,clip]{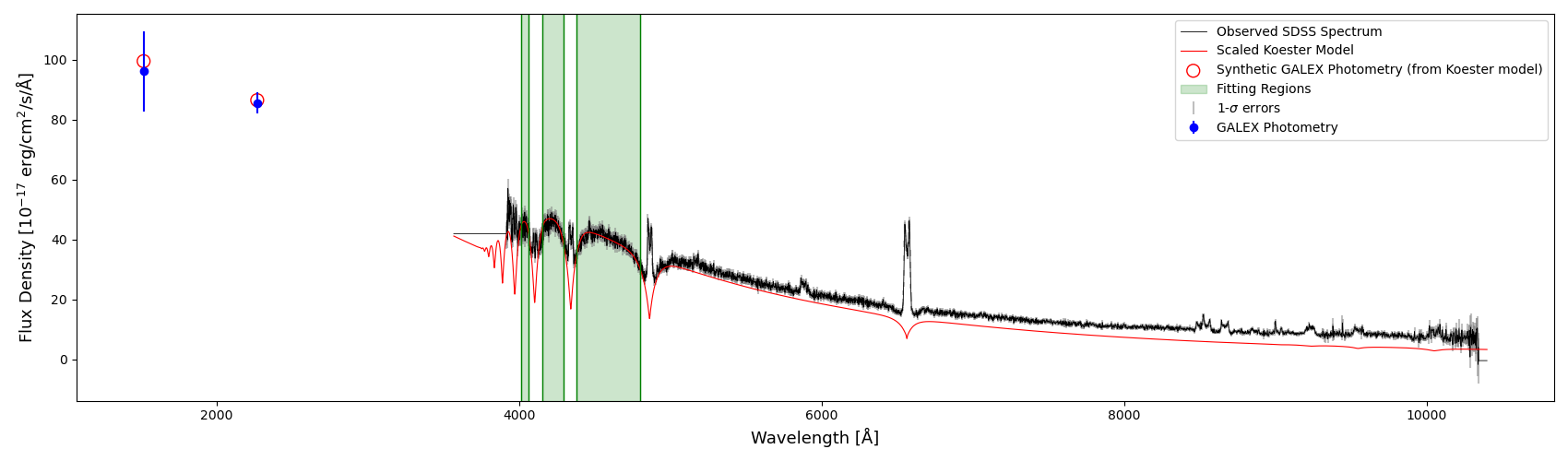}\hfill
    \includegraphics[width=\textwidth,trim=0.4cm 0.3cm 0.3cm 0.2cm,clip]{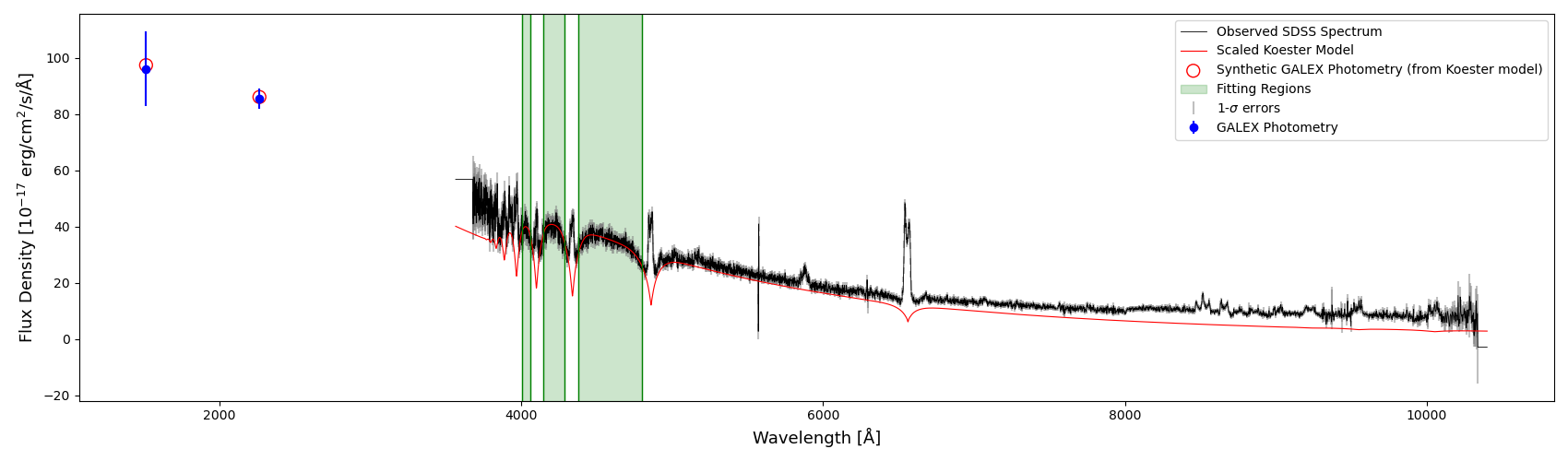}
    \caption{SDSS spectra of GALEX J125751.4-283015 fitted with a Koester WD model. The GALEX photometry is used to better constrain the temperature. \textit{Upper panel:} First spectrum with the best fit model characterized by $\log(g)$=8.0 and T$_{\rm eff}$=12000\,K. \textit{Lower panel:} Second spectrum with the best fit model characterized by $\log(g)$=8.5 and T$_{\rm eff}$=12500\,K.}
    \label{GALEX1257_SDSSfit}
\end{figure*}

For GALEX J125751.4-283015 we have two SDSS spectra available (see Fig.~\ref{GALEX1257_SDSSfit}) that we complement with GALEX photometry. In the following, we describe the procedure used to determine the effective temperature and surface gravity of the WD.

In isolated WDs, the Balmer absorption lines are typically used to constrain the effective temperature and surface gravity by fitting a grid of WDs atmosphere models to the observed spectra. In accreting WDs, this procedure becomes more complicated due to superimposed Balmer emission lines originating from the accretion disk as well as additional continuum contributions from the accretion disk and the donor. In CVs, the WD typically maintains an effective temperature of $T_{\text{eff}}$ $\gtrsim10000$\,K, due to compressional heating from the accreted material,
thus, dominating in the blue region of the optical spectrum (see e.g. \citealt{Sion_1995} and \citealt{Sion_1999}). Continuum emission from the disk becomes increasingly important at wavelengths $\gtrsim5000$\,\AA, while the donor in low mass-transfer rate systems, such as period-bouncers, is of very low mass and starts to contribute only in the near–infrared regime, that is $\lambda\gtrsim7000$\,\AA.

We used a grid of pure-hydrogen, Local Thermodynamic Equilibrium (LTE) WD atmosphere models (\citealt{koester2010}) covering $T_{\rm {eff}}=8000-20000$\,K in steps of 250\,K and $\log(g)$=6.5 - 9.5 in steps of 0.25. The WD models were interpolated to the wavelengths of the observed spectrum using a cubic spline. To match the spectral resolution of the SDSS data, the WD models were then convolved with the Gaussian instrumental profiles of the SDSS observed spectra. To further constrain $T_{\rm{eff}}$, we additionally use GALEX near- and far-ultraviolet photometry (obtained as detailed in \citealt{munoz2024}). The ultraviolet flux is highly sensitive to $T_{\rm{eff}}$, making it particularly valuable for breaking degeneracies in the optical fitting. For the comparison, synthetic photometric fluxes were computed from the WD models using the GALEX near- and far-ultraviolet response functions (see \citealt{Casagrande}). To fit the WD model to an observed spectrum, we minimized a total chi-squared function which combines the spectroscopic component—evaluated over selected wavelength intervals—with the photometric components from GALEX,

\begin{align}
& \chi^2 = \sum_{i} \left( \frac{F_{\text{obs},i}(\lambda) - K_{\text{WD}} F_{\text{model},i}(\lambda)}{\sigma_{\text{obs},i}(\lambda)}\right)^2\\
& +\sum_{j} \left( \frac{F_{\text{obs, GALEX},j} - K_{\text{WD}} F_{\text{model, GALEX},j}}{\sigma_{\text{obs, GALEX},j}}\right)^2,
\label{eq:chi_squared}
\end{align}

where $F_{\text{obs},i}$ and $\sigma_{\text{obs},i}$ are the observed flux values and their uncertainties and $F_{\text{model},i}$ are the corresponding flux values from the WD model. The terms $F_{\text{obs, GALEX},j}$ and $\sigma_{\text{obs, GALEX},j}$ represent the observed GALEX fluxes and their uncertainties in the far- and near-ultraviolet bands, while $F_{\text{model, GALEX},j}$ are the synthetic fluxes obtained from the WD model. The scaling factor $K_{\text{WD}}$ is the free parameter to optimize.

The fit of the spectroscopic component is performed over three wavelength intervals: [4010, 4060], [4150, 4290], and [4380, 4800]\,\AA (see Fig.~\ref{GALEX1257_SDSSfit}). These intervals are selected to exclude the red optical regime where the accretion disk or the donor may contribute producing excess in the continuum flux. These wavelength intervals also mask the Balmer emission cores from the accretion disk, while capturing the broad wings of the Balmer absorption lines from the WD, helping to constrain $\log(g)$. The final best-fit model is selected from the models with the lowest total chi-squared values (Eq.~\ref{eq:chi_squared}), by choosing the one that shows the best agreement with the GALEX photometry.


We fitted the two SDSS spectra of GALEX\,J125751.4-283015 using the methodology described above obtaining a $\log(g)$ between $8.0$ and $8.5$ and a WD effective temperature between $12000$\,K and $12500$\,K, values typical of short period CVs and known period-bouncers. Once the best fitting model was established, we can use it to estimate the WD radius and mass. Hereby, we considered that the ratio between the observed flux from the SDSS spectrum and the model surface flux represents the dilution factor, $(d/R_*)^2$, where $d$ is the distance and $R_*$ is the stellar radius. The two SDSS spectra of GALEX J125751.4-283015 and the distance of 180\,pc reported by \cite{bailer2021} give a WD radius between 8.01$\times 10^8$\,cm and 8.93$\times 10^8$\,cm. From these values, we derive the WD mass using the mass-radius relation by \cite{nauenberg}, finding 0.67M$_{\odot}$ and 0.57M$_{\odot}$, from the two spectra, respectively. The mass range that we obtain for the WD is below the average value for CVs ($\approx 0.8\,{\rm M_{\odot}}$; \citealt{pala2020}, \citealt{pala2022}) and would represent one of the period-bouncers with the lowest WD mass, together with V379 Vir with a WD mass of 0.64M$_{\odot}$ \citep{munoz2023}.

\end{appendix}

\end{document}